\documentclass[aps,prl,epsf,twocolumn]{revtex4}
\usepackage{graphicx,epsf}
\usepackage{dcolumn}
\usepackage{bm}

\begin{document}


\def\appConv{Appendix~A}
\def\appBeer{Appendix~B}
\def\appHValid{Appendix~C}

\def\co2{{CO$_2$}}
\def\h2o{{H$_2$O}}
\def\n2{{N$_2$}}
\def\cubtc{Cu-BTC}
\def\ow{{O$_W$}}
\def\HfO2{{HfO$_2$}}
\def\SiO2{{SiO$_2$}}
\def\Al2O3{{Al$_2$O$_3$}}
\def\k{{$\kappa$~}}
\def\ks{{$\kappa_s$~}}
\def\ksten{{$\tensor{\kappa}_s$~}}
\def\kinf{{$\kappa_\infty$~}}
\def\kion{{$\kappa_{ion}$~}}
\def\kionten{{$\tensor{\kappa}_{ion}$~}}
\def\kinften{{$\tensor{\kappa}_\infty$~}}
\def\cm1{{cm$^{-1}$}}
\def\zstar{{$Z^{\star}$~}}

\def\comment#1{{\large\textsl{#1}}}
\def\degree {{$^\circ$}}
\def\degrees{{$^\circ$}}
\def\eq#1{{Eq.~(\ref{eq:#1})}}
\def\fig#1{{Fig.~\ref{fig:#1}}}
\def\sec#1{{Sec.~\ref{sec:#1}}}
\def\inv{^{-1}}
\def\micron {\hbox{$\mu$m}}
\def\microns{\micron}
\def\Ref#1{{Ref.~\onlinecite{#1}}}  
\def\tab#1{{Table~\ref{tab:#1}}}
\def\tauOne{\tau^{(1)}}
\def\tVec{\hbox{\bf t}}
\def\thetaDet{\theta_{DET}}

\def\qvec{{\vec q}}
\def\pvec{{\vec p}}
\def\Avec{{\vec A}}
\def\qhat{{\hat q}}
\def\qperphat{{\hat q_\perp}}
\def\ekpq{{E_{\kvec+\qvec}}}
\def\ek{{E_{\kvec}}}
\def\Omegabar{{\bar\Omega}}
\def\omegabar{{\bar\omega}}
\def\omegap{{\omega_p}}
\def\kf{{k_F}}
\def\kappaf{{\kappa_F}}
\def\mone{{-1}}
\def\re{{\rm{Re\,}}}
\def\im{{\rm{Im\,}}}
\def\twopi{{2 \pi}}
\def\wpm{w_\pm}
\def\FWlindhard{Appendix~A}
\def\lindhardTrans{Appendix~B}
\def\ftr{{f^{tr}}}
\def\PN{Pines and Nozi{\`e}res}
\def\Bohm{B{\"o}hm}
\def\Nifosi{Nifos{\'\i}}
\def\prin{{\cal P}}

\def\imagOmegaSq{the Appendix}
\def\epsTensor{{\buildrel \leftrightarrow \over \epsilon}}
\def\chiTensor{{\buildrel \leftrightarrow \over \chi}}
\def\idenTensor{{\buildrel \leftrightarrow \over I}}
\def\epsTrans{{\epsilon^{(t)}}}
\def\epsLong{{\epsilon^{(\ell)}}}
\def\epsTransInv{{\epsilon^{(t)-1}}}
\def\epsLongInv{{\epsilon^{(\ell)-1}}}

\def\MvecA{{M^{(\vec A)}}}
\def\MdivA{{M^{(\nabla \cdot \vec A)}}}
\def\Mphi{{M^{(\phi)}}}
\def\backGrad{{\buildrel \leftarrow \over \nabla}}
\def\backMom{i \hbar \backGrad}

\def\half{{1/2}}
\def\minusHalf{{-1/2}}
\def\threeHalves{{3/2}}
\def\minusThreeHalves{{-3/2}}

\newenvironment{bulletList}{\begin{list}{$\bullet$}{}}{\end{list}}


\title{Density Functional Theory Meta-GGA+U Study of Water
Incorporation in the Metal Organic Framework Material
Cu-BTC}

\author{Eric Cockayne\footnote{Electronic address: eric.cockayne@nist.gov}}

\affiliation{Materials Measurement Science Division, 
Material Measurement Laboratory, 
National Institute of Standards and Technology, 
Gaithersburg, Maryland 20899 USA}

\author{Eric B. Nelson\footnote{Current address: Materials Science and
Engineering Department, Boise State University, Boise, ID 83725 USA}}

\affiliation{Materials Measurement Science Division,
Material Measurement Laboratory,
National Institute of Standards and Technology,
Gaithersburg, Maryland 20899 USA}


\date{\today}

\begin{abstract}

 Water absorption in the metal-organic framework (MOF) material
Cu-BTC, up to a concentration of 3.5 H$_2$O per Cu ion,
is studied via density functional theory at the meta-GGA+U
level.  
The stable arrangements 
of water molecules show chains of hydrogen-bonded water molecules 
and a tendency to form closed cages at high concentration.
Water clusters are stabilized primarily by a combination of water-water
hydrogen bonding and Cu-water oxygen interactions.
Stability is further enhanced by van der Waals interactions, 
electric field enhancement of water-water bonding, and hydrogen 
bonding of water to framework oxygens.  We hypothesize that 
the tendency to form such stable clusters explains the particularly 
strong affinity of water to Cu-BTC and related MOFs with 
exposed metal sites.


\end{abstract}


\maketitle
\thispagestyle{empty}

\section*{Introduction}
\label{sec:intro}

 The need to reduce greenhouse gas emissions has driven the
search for materials to capture and sequester 
carbon dioxide at a low-enough cost to  be economically viable.
Metal-organic framework (MOF) materials
offer great promise for the capture of 
CO$_2$.\cite{Millward05,Sumida11}
Absorption and desorption from MOF materials via swings of
temperature and pressure in principle requires less energy than
do current aqueous capture technologies.\cite{Sumida11}
In addition, MOF materials can be created with a wide variety
of pore sizes, shapes, connectivities, and topologies.
These attributes, as well as the effects of cation substitution 
on metal sites, 
can in principle be used to tailor both the physical 
and chemical interactions of the MOF with gas molecules to 
optimize desirable properties such as CO$_2$ selectivity, capture rates, 
and costs of absorption/desorption cycles.\cite{Millward05}

\begin{figure}
\includegraphics[width=236pt]{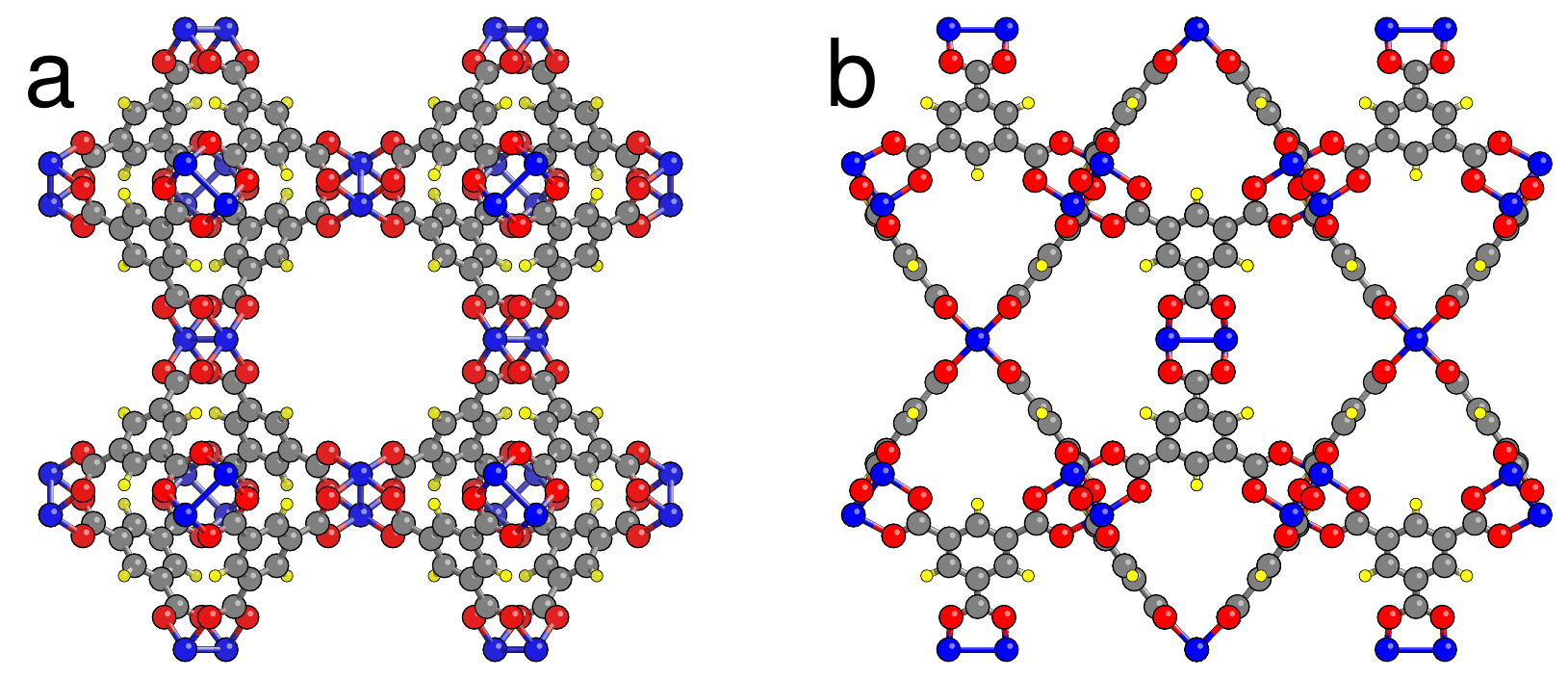}
\caption{Cu-BTC viewed along (a) 100 and (b) 110 directions. Cu atoms
are blue; O red; C gray; H small yellow.}
\label{fig:struct}
\end{figure}

A particularly interesting and well-studied example of a MOF material is \cubtc, 
also known as HKUST-1 (Ref.~\onlinecite{Chui99}).
This material (\fig{struct}) consists of copper dimers linked by
1,3,5-benzenetricarboxylate C$_6$O$_9$H$_3$ (BTC) units.
The structure has a three-dimensional cubic framework with
channels of alternating 13.3 \AA~and 11.1 \AA~cuboctahedral pores connected
by 6.4 \AA~square windows along 100-type directions, and 
5.5 \AA~tetrahedral side pockets connected to the 13.3 \AA~pores via 
3.7 \AA~triangular windows.\cite{Vishny03,foot1}

As formed in atmospheric conditions, \cubtc~ contains a significant
degree of water.
This water can be removed via heat-treatment under near 
vacuum,\cite{Schlichte04,Chui99,Grzech11} leaving a structure with 
partially ``exposed" Cu ions\cite{Brown09} facing the large
cuboctahedral pores, allowing for 
particularly strong interactions with adsorbates at these Cu sites.
These exposed ions lead to several potential applications in addition
to carbon capture, including catalysis,\cite{Schlichte04}
hydrogen storage,\cite{Peterson06} storage of other gases such as 
NO,\cite{Xiao07} and gas separation.\cite{Yang06}

 Dry \cubtc~ has a \co2 uptake of as much as 19.8 weight~\%
at atmospheric pressure,\cite{Aprea10} and a high selectivity
of \co2 over \n2.\cite{Aprea10,Sumida11}
The situation of hydrated \cubtc~ is interesting.  Up to about
one \h2o per Cu site, there is theoretical and experimental
evidence for a slight increase in \co2 uptake.\cite{Yazaydin09,Liu10,Yu12}
For water concentrations greater than about one \h2o per Cu, a situation
which we call here ``highly-hydrated \cubtc",
the \co2 uptake is reduced.\cite{Liu10,Chowdhury12,Yu13}  At high-enough concentrations
of \h2o, \cubtc~loses almost all of its \co2 capacity.\cite{Liu10}
The last fact likely prevents the use of \cubtc~as a material for
post-combustion carbon capture, as flue gases in coal-burning
plants contain significant water vapor.\cite{Espinal12}
In some MOFs, water can even break down the structure completely.\cite{Schoenecker13}

In spite of the importance of the highly-hydrated state of \cubtc~for its performance,
little is known about the structure of water in this state.
Absorption isotherm measurements\cite{Liu10,Schoenecker13} show that as much as 32
to 40 mol kg$^{-1}$ of water can be absorbed into \cubtc.  Assuming an ideal 
pore structure, this corresponds to 6.5 to 8.0 \h2o per Cu.
An NMR study\cite{GulENoor13} suggests that the state of highly-hydrated
\cubtc~ has one  water molecule bound to each Cu in equilibrium with fluid water 
in the remaining pore volume.
A recent X-ray powder diffraction refinement,\cite{Winnie14} on the other hand,
shows evidence for 2.3 bound water molecules per Cu atom, in three different partially
occupied binding sites, although it was not possible to determine the 
hydrogen positions for these water molecules.

 Previous electronic structure calculations and molecular simulations have generally 
investigated the structure of \h2o in \cubtc~ only  up to one 
\h2o per Cu ion,\cite{Castillo08,Grajciar10,Supro13,Toda13} in which case  
one water molecule binds to each Cu position, with the water oxygen atom
(\ow) closest to the Cu. In this work, we use density functional theory at
the ``meta-GGA + U" level to investigate the structure of water absorption
in Cu-BTC up to 3.5 \h2o per Cu.

\section*{Computational Methods}

First principles density functional theory calculations, as encoded in
the {\sc VASP} software~(\onlinecite{Kresse96,disclaim}), were used to calculate the 
relaxed configurations investigated here and their electronic 
structures.  
All calculations were performed for a primitive cell of 
\cubtc~containing 156 framework atoms plus any H$_2$O adsorbates.
Because of the large cell, only a single k-point at the origin was used.
The planewave cutoff was 500 eV for all calculations.
Van der Waals forces were treated using the ``DFT-D2" approximation 
of Grimme,\cite{Grimme06}
and were included in all calculations.

\begin{table} \caption{Total binding energies of water dimers
and hexamers (in eV). {\textsc VASP} results are compared with benchmark
{\textsc GAUSSIAN} results.
The {\textsc GAUSSIAN} results show the average and (in parenthesis, with
units of the least significant digit) root mean square deviation for 
three different highly-converged basis sets,\cite{foot2} as tabulated in 
Ref.~\onlinecite{Lee00}.
Only the electronic contribution to binding energy is included.}
 \begin{tabular}{ccc}
  \hline
 DFT code & water dimer   & water hexamer \\
{\textsc VASP} GGA      & -0.30    & -2.92  \\
{\textsc VASP} meta-GGA & -0.23    & -2.13   \\
{\textsc GAUSSIAN}      & -0.23(1) & -2.05(16) \\
\hline
\label{tab:vvsg}
\end{tabular} \end{table}

 To study water absorption in \cubtc, it is crucial to get the 
interactions between \h2o molecules correct.  We compared various
exchange-correlation (XC) functionals to see if the correct binding
energies of small water clusters could be obtained.  The results
for two such XC, the PBEsol\cite{Perdew08} generalized gradient 
approximation (GGA), and the PBEsol plus RTPSS\cite{Sun11} meta-GGA functional 
(one that also uses the {\em second} derivative of the charge density), 
are shown in \tab{vvsg}, and compared
with benchmark calculations of water binding energies\cite{Lee00} that were performed
using the DFT code {\textsc GAUSSIAN}\cite{disclaim} 
with highly-converged basis sets.\cite{foot2}
The results show 30~\%~to 40~\%~ water-water overbonding for the
case of the GGA calculations, but excellent agreement with the
{\textsc GAUSSIAN} results for the meta-GGA approximation.  
The meta-GGA approximation was therefore used for the rest of this work,
despite being several times more computationally expensive than the GGA approximation.

 As noted in previous DFT works, improvement in the agreement
between DFT results and experiments for systems containing
magnetic ions can generally be achieved if on-site Coulomb terms are
included,\cite{Liech95} in what is commonly termed the ``GGA + U" approach.
To determine the onsite Coulomb parameters to use for \cubtc, we began
with the experimental structure\cite{Zigan77,Merlini12} of Cu$_2$CO$_3$(OH)$_2$ (malachite).
Malachite has numerous structural features
analogous to those in \cubtc, including the same species 
(Cu,O,C,H), Cu in the Cu$^{2+}$ valence state, Cu coordinated to a square of oxygen, 
and OH units which are analogous to the H$_2$O admolecules in hydrated \cubtc, but
malachite
is simpler to investigate because of its smaller unit cell.

 Our aim was to adjust the U parameter for both Cu and O so as
to fit the bandgap and experimental crystallographic structure of malachite 
as well as possible, and, assuming transferability, to use the same 
values to study \cubtc.
We were unable to find any value of the malachite bandgap in the literature.
Our calculations, however, suggested that a
bandgap of 1.75 eV leads to a minimum of absorption at around 2.3 eV, 
at the characteristic green color of malachite, and we thus fit to this
value.
We used the malachite coordinates given by Zigan {\it et al.}
(Ref.~\onlinecite{Zigan77}, cited in Ref.~\onlinecite{Girgsdies12}), because
hydrogen positions were given.
To quantify the structural agreement with experiment, 
we fixed ions at their experimental positions and calculated the root 
mean square residual forces on the ions.
For a calculated bandgap of 1.75 eV, the residual forces were 
minimized for Cu U = 3.08 eV and oxygen U = 7.05 eV.
The magnitude of the  U value for Cu largely controls the 
calculated bandgap, by (controlling the splitting in the Cu d levels),
while including a nonzero U for oxygen greatly reduces the residual forces.
Magnetism of Cu$^{2+}$ ions with their $d^9$ electronic
configurations was treated using spin-polarized DFT
calculations.  The antiferromagnetic arrangement of the 2 Cu
on the Cu-Cu dimer was found to be lower in energy than the ferromagnetic one, 
in agreement with previous studies.\cite{Poeppl08}

\begin{figure}
\includegraphics[width=236pt]{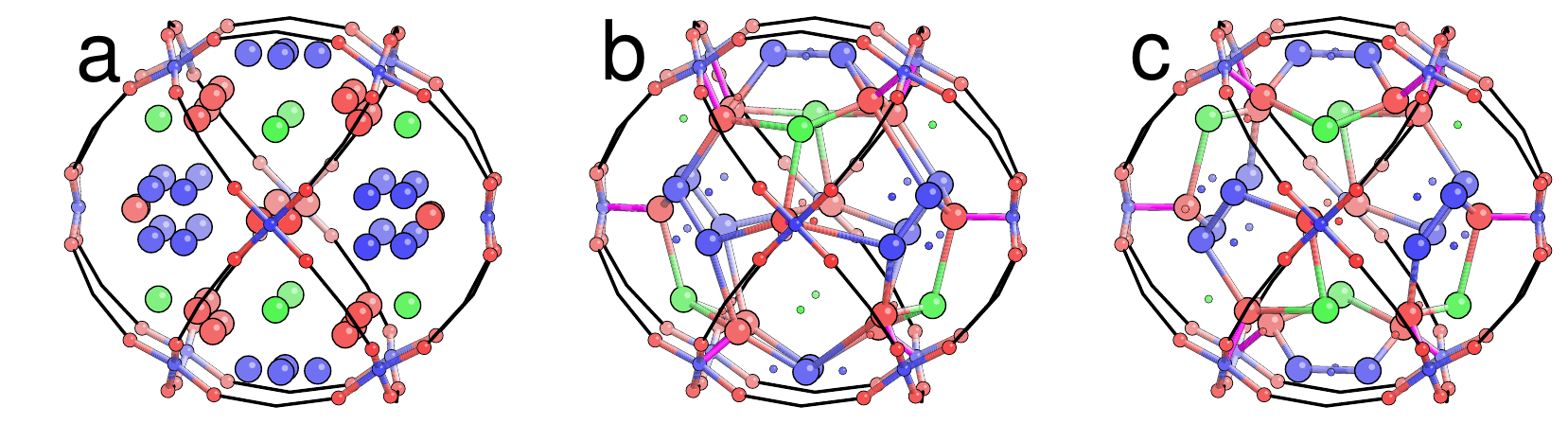}
\caption{(a) Large spheres show positions of water oxygens (\ow) inside the
large pore of \cubtc, according to the
structure refinement of Wong-Ng {\it et al.}\cite{Winnie14} 
Sites are color-coded according to symmetry (green: \ow(I); blue: \ow(II); 
red: \ow(III)).  Experiment shows partial occupancy for each type of \ow~site;  
corresponding to an average of about 28 of the 56 \ow~sites shown occupied
per large pore.  (b) Way of occupying 28 \ow~sites of (a) without any pair
of water molecules approaching too close.  Cu-\ow(III) bonds shown in magenta.
(c) Way of occupying 30 \ow-sites of (a).}
\label{fig:deco28}
\end{figure}

 We now hydrate the \cubtc~structure.  We begin with the
positions of the water molecule oxygens (\ow) determined by
Wong-Ng {\it et al.}\cite{Winnie14} (\fig{deco28}; \tab{single}), all
located inside the large 13 \AA~pores.  In this study, 
the \ow(I) sites (green in \fig{deco28}) were found to have an occupancy 
of 0.80(2), the \ow(II) sites (blue) occupancy 0.32(1) and the \ow(III) site 
(red) occupancy 0.57(2).
The partial occupancy values imply an average of about 28 \h2o~molecules per
large  pore in arrangements that vary between pores.
Physically, the partial occupancies are mandated because
many of the potential sites are too close together
to be simultaneously occupied.  In \fig{deco28}(b) and \fig{deco28}(c),
we show two different choices for placing \ow~that avoid unphysically
close pairs of water molecules.
The first has an arrangement of 28 water molecules with the 
\ow~forming a polyhedra, and
the second consists of 6 independent cyclic clusters of 5 water molecules,
for 30 total water molecules.  We refer to models with these starting 
configurations of \ow~as ``Model-28" and ``Model-30",
respectively, where the number refers to how many water molecules
per primitive \cubtc~cell, or, equivalently, per large pore.

\begin{figure}
\includegraphics[width=236pt]{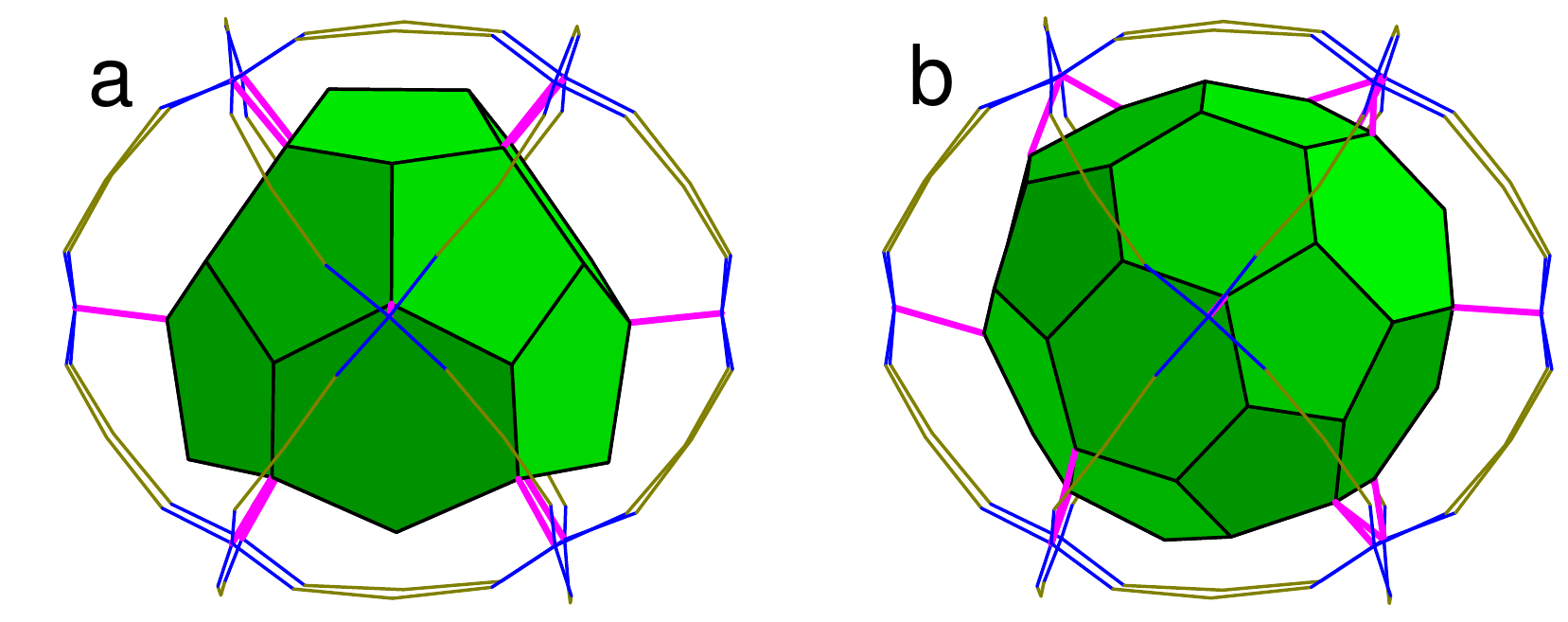}
\caption{Topology of water oxygens (\ow) in two different models for water
structure of highly-hydrated \cubtc. (a) Model-28, corresponding to the
\ow~network in Fig. 2(b). (b) Model-42, a fullerene-like arrangement
of 42 \ow.}
\label{fig:bucky}
\end{figure}

 The choice in \fig{deco28}(c) has the same topology as
the C$_{28}$ fullerene of T$_d$ symmetry.\cite{Atlas}
We show the topology in \fig{bucky}(a).
In fullerenes, each carbon has three carbon neighbors.  In the
equivalent water structure, a highly hydrogen-bonded structure
is produced in an analogous way to the ice rules for three-dimensional
ice: each \ow~is covalently bonded to two hydrogens, and each 
\ow-\ow~neighbor link has one hydrogen atom that is bonded to one
\ow~and forms a hydrogen bond to the other.  
These observations inspired us to look for other fullerene-like
arrangements of \ow~molecules that might be accommodated in \cubtc.
We looked at all low-energy fullerene geometries in the online Atlas of
Fullerene Structures\cite{Atlas} from $N$ = 20 to $N$ = 60 at various
orientations and scalings within the \cubtc~large pore to see what structure
would best accommodate preferred \ow-\ow~distances of 2.9 \AA,
while having one \ow~near each exposed Cu.   The best candidate
structure is a 42-molecule structure equivalent to the
``C$_{42}$ \#45" fullerene, where the \ow~have the topology
shown in \fig{bucky}(b).  We call this ``Model-42".
 For each model, as well as a model ``Model-12", with one
\h2o per Cu, we initially randomized the hydrogen positions,
then performed a full relaxation until all forces were 
converged within 0.03 eV \AA$^{-1}$.


\section*{Results and Discussion}

 The calculated structure of dry \cubtc~is shown in~\tab{dry},
and compared with a recent experimental X-ray powder diffraction
refinement.\cite{Winnie14}
Extremely good agreement (within 0.02~\AA) is obtained for all atomic positions
except for H, which are 0.14~\AA~from the experimental positions.
We note, however, that the experimental refinement treated the
organic ligand as a rigid body; and did not further refine the average hydrogen 
positions, which have little effect on the X-ray powder diffraction 
pattern in any case.

\begin{table} \caption{Comparison of symmetrized DFT and experimental\cite{Winnie14} 
structures of dry \cubtc. Space group Fm$\overline{3}$m with
$a$ = 26.2793~\AA.  The standard deviation of the experimental structure
refinement coordinates are 0.0001 or smaller.}
 \begin{tabular}{cccccccccc}
  \hline
     &              &      & DFT  &   & &   &   Expt. &   \\
Atom & Wyckoff pos. & x    & y     & z & & x & y & z  \\
Cu   & 48(h)  & 0.0000 & 0.2170 &  0.2170 & & 0.0000 & 0.2164 & 0.2164 \\
O    & 192(l) & 0.1833 & 0.2437 &  0.4478 & & 0.1831 & 0.2438 & 0.4480    \\
C(1) & 96(k)  & 0.2035 & 0.2035 &  0.4308 & & 0.2037 & 0.2037 & 0.4302 \\
C(2) & 96(k)  & 0.1786 & 0.1786 &  0.3864 & & 0.1783 & 0.1783 & 0.3865 \\
C(3) & 96(k)  & 0.1352 & 0.1352 &  0.2996 & & 0.1352 & 0.1352 & 0.3000 \\
H    & 96(k)  & 0.1189 & 0.1189 &  0.2654 & & 0.1202 & 0.1202 & 0.2703 \\
\hline
\label{tab:dry}
\end{tabular} \end{table}


As a bridge to the water cluster studies, we
first investigated the binding energy of a single water molecule
in \cubtc.
We tested each \ow~position found in the Wong-Ng {\it al.} (W-N) powder 
diffraction structure refinement.\cite{Winnie14}  For comparison, we also tested 
\ow~in the high-symmetry locations in \cubtc: the centers of the large, medium, 
and small pores, and the centers  of the square and triangular windows 
between pores.  The center of the small tetrahedral pore has been
found to be a favored bonding site for certain sorbates in
previous work on \cubtc.\cite{Castillo08,
Hulvey13}
We additionally investigated a medium pore ``bonding", 
site set 3.8 \AA~from the center of the C$_6$ ring, chosen because it is
a favorable distance for water-carbon separation.\cite{Ambrosetti11}
For simplicity, the \cubtc~framework was kept rigid in each case and
only the hydrogens were allowed to move, except in the case of
W-N \ow(III), where, in addition, a full relaxation
was performed.
We broke down the total binding energies $\Delta{E}$ into two
parts: molecule-framework van der Waals energy (vdW), and the
excess part, due to chemical (bonding/antibonding  and nondispersive
electrostatic) interactions of the molecule with the framework (chem).
The vdW (dispersion) energy
was calculated using the approximation of Grimme, only counting
vdW interactions between the \h2o~molecule and the framework.
The chemical component of the binding energy was
obtained by simply subtracting the vdW contribution from the total.
The results are shown in~\tab{single}.


\begin{table}
\caption{Calculated binding energies $\Delta{E}$ of a single water molecule in~\cubtc. 
at various positions of the \ow. ``W-N" sites are from Ref.~\onlinecite{Winnie14}.
Binding energies are broken down into van der Waals (vdW) and
chemical (chem) contributions. \ow~positions are given in crystallographic notation and
all energies are in eV.}
\begin{tabular}{ccccccc}
  \hline
\ow~position                         &  x     &  y     &   z    &  $\Delta{E}$   &  vdW & chem  \\
W-N I                                & 0.1307 & 0.1307 & 0.1307 & -0.14        & -0.08 & -0.06 \\
W-N II                               & 0.2287 & 0.0364 & 0.0364 & -0.12        & -0.06 & -0.06 \\
W-N III                              & 0.1696 & 0.1390 & 0.0000 & -0.47        & -0.12 & -0.35 \\
W-N III + full relaxation            & 0.1712 & 0.1413 & 0.0000 & -0.53        & -0.13 & -0.40 \\
Large pore center                    & 0.0000 & 0.0000 & 0.0000 & -0.01        & -0.01 & -0.00 \\
Medium pore center                   & 0.5000 & 0.5000 & 0.5000 & -0.02        & -0.01 & -0.01 \\
Small pore center                    & 0.2500 & 0.2500 & 0.2500 & -0.12        & -0.09 & -0.03 \\
Square window                        & 0.0000 & 0.0000 & 0.2564 & -0.05        & -0.03 & -0.02 \\
Triangular window                    & 0.1900 & 0.1900 & 0.1900 & -0.18        & -0.14 & -0.04 \\
Medium pore ``bonding"               & 0.4260 & 0.4260 & 0.4260 & -0.06        & -0.04 & -0.01 \\
\hline
\label{tab:single}
\end{tabular} \end{table}

As expected water binds most strongly to the W-N III site (the
one near the exposed Cu ion) and this binding is mostly due to chemical interactions. 
The second-strongest binding site found is the triangular window, due to large van der Waals interactions, 
but this site is absent from the experimental structure refinements for some reason.  
The small pore center and W-N I and W-N II sites bind water molecules in the -0.12 eV to -0.14 eV range.  
Aside from the W-N III site, sites W-N I and W-N II have the strongest chemical interactions.
Bonding is weak at the large and medium pore centers due to distance from the framework and bonding
is weaker in the medium pore than the large and small pores due mainly to the lack of chemical interactions
with the framework.

\begin{figure}
\includegraphics[width=236pt]{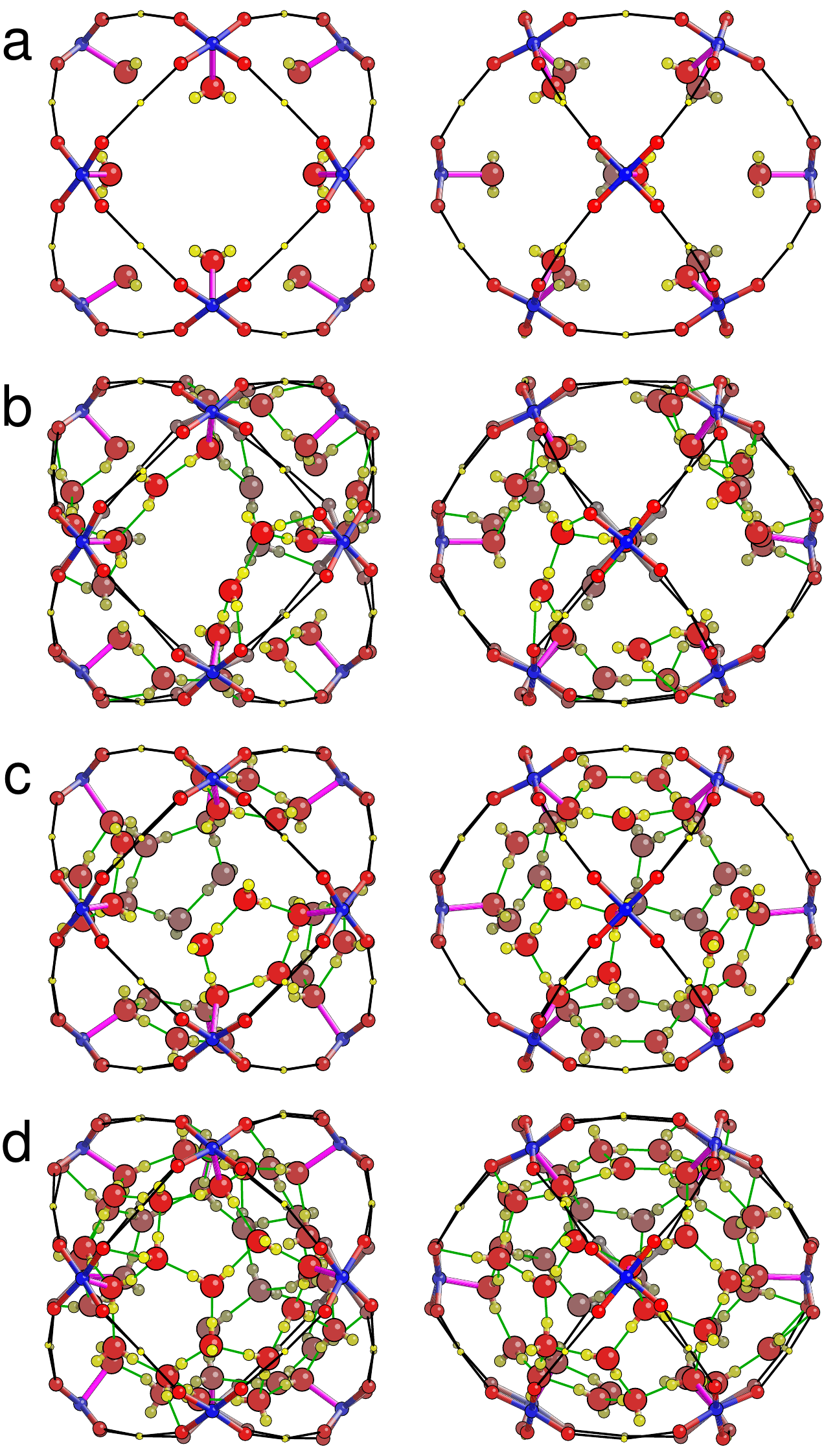}
\caption{Structures of model water clusters inside large pore of \cubtc,
relaxed via density functional theory.
(a) Model-12; (b) Model-28; (c) Model-30; (d) Model-42, where
the number refers to the number of water molecules inside the large pore.
Hydrogen bonds are shown by thin green lines.}
\label{fig:relaxed}
\end{figure}

\begin{table} 
\caption{Key results for the different models for hydrated~\cubtc~
investigated with this work, compared with experimental data (where available).
Distances $D$ are in \AA~and energy differences $\Delta{E}$ are in eV.
Standard deviations of the experimental structure refinement results are given
(in parentheses, with units of the least-significant digit).}

\begin{tabular}{ccccccc}
  \hline 
Study           & $N_W$ & D$_{Cu-Cu}$ & D$_{Cu-O}$ & D$_{Cu-O_W}$ & $\Delta{E}$ & $\Delta{E}$/$N_W$  \\
DFT  (dry)       & 0     &  2.454      & 1.932     &             &                   &                  \\
DFT Model-12     & 12    &  2.537      & 1.946     &  2.262      &  -6.30            &   -0.52          \\
DFT Model-28     & 28    &  2.582      & 1.954     &  2.241      &  -16.66           &   -0.59          \\
DFT Model-30     & 30    &  2.589      & 1.950     &  2.263      &  -17.49           &   -0.58          \\  
DFT Model-42     & 42    &  2.586      & 1.954     &  2.217      &  -26.81           &   -0.64          \\
Expt. (dry)      &  0    &  2.498(1)   & 1.930(1)  &             &                   &                  \\
Expt. (hydrated) & 27.84 &  2.628(10)  & 1.951(6)  &  2.32(2)    &                   &                  \\
\hline 
\label{tab:key}
\end{tabular} \end{table}





Now we give the results for \cubtc~with multiple water molecules.
The relaxed structures for the hydrated \cubtc~models are shown in \fig{relaxed}.
Hydrogen bonds are shown as thin green lines.
The key structural parameters and energetics results are shown in~\tab{key}.
As the number of water molecules increases, the
structure changes from individual water molecules
bound to Cu to clusters of water molecules bound
to one or more Cu, to a closed cage encompassing
all of the water molecules.
The trend in key interatomic distances with increasing
numbers of water molecules agrees with
experiment.

Although Model-28 and Model-30 begin with
atoms at positions suggested by experimental structure refinement,
certain water oxygen positions relax as much as 1.6~\AA,
suggesting a significant (quasi)static contribution to the 
large experimental displacement factors found for water oxygens.\cite{Winnie14}
The relaxed DFT positions do remain
within and near the inner surface of the
large 13 \AA~pore.  Our Model-42 results show that 
3.5 molecules per Cu are easily accommodated in this
inner surface region.  If water also occupies the interior of the
large pore as well as the 11 \AA~and 5.5 \AA~pores, then
\cubtc~can accommodate the 6.5 to 8.0 \h2o~molecules
per Cu ion suggested by experiment without requiring
larger (defect) pores.

\begin{table*}
\caption{Breakdown of interactions of water clusters in \cubtc.  Total
cluster binding energy $\Delta{E}$ and cluster binding energy per water molecule
($\Delta{E}/N_W$ are broken down into three components: intracluster (intra), 
cluster-framework Van der Waals (vdW), and cluster-framework chemical interactions (chem). 
All energies in eV.}
 \begin{tabular}{ccccccccccc}
  \hline
Model            & $N_W$ & $\Delta{E}$  &          &        &       & $\Delta{E}/N_W$ &   &  &       \\
                 &       &  total       &  intra   &  vdW  &  chem  & total & intra &  vdW  &  chem \\
DFT Model-12     & 12    &  -6.30       &  -0.22   & -1.49 &  -4.59 & -0.52 & -0.02 & -0.12 & -0.38   \\
DFT Model-28     & 28    &  -16.66      &  -8.13   & -2.66 &  -5.86 & -0.59 & -0.29 & -0.10 & -0.21   \\
DFT Model-30     & 30    &  -17.49      &  -9.96   & -2.60 &  -4.93 & -0.58 & -0.33 & -0.09 & -0.16   \\
DFT Model-42     & 42    &  -26.81      & -16.64   & -3.91 &  -6.25 & -0.64 & -0.40 & -0.09 & -0.15   \\
DFT Model-28MP   & 28    &  -14.81      & -12.72   & -2.10 &   0.01 & -0.53 & -0.45 & -0.07 &  0.00   \\
\hline
\label{tab:break}
\end{tabular} \end{table*}

\begin{figure}
\includegraphics[width=236pt]{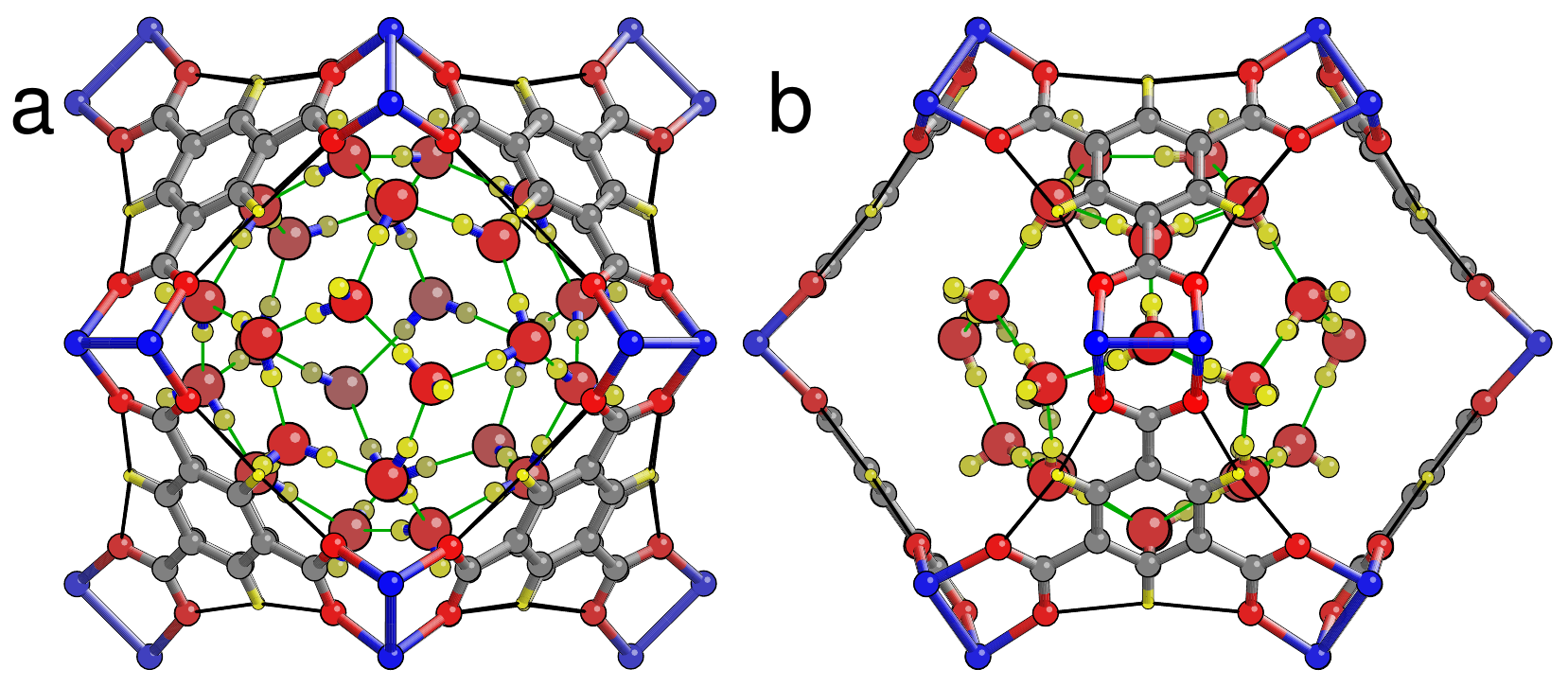}
\caption{Structure of a 28-water cluster inside a medium pore
of \cubtc, relaxed via density functional theory.
(a) 100 view  (b) 110 view.}
\label{fig:medium}
\end{figure}

The binding energy per water molecule for one \h2o~attached to
each Cu site is -0.52~eV, almost identical to that for an isolated 
\h2o~molecule (\tab{single} {\it vs.}~\tab{key}).   
In the highly-hydrated state configurations, the magnitude of the 
binding energy per water molecule is even larger.  Where does this enhanced
stability come from?
To determine this, we broke down the binding energy of each model
into three parts: intracluster energy; cluster-framework van der
Waals energy (vdW), and the excess part, due to chemical 
interaction of the cluster with the framework.  The intracluster binding
energy was calculated by removing the framework and performing a DFT
energy calculation on just the cluster.  
The remaining contribution to the binding energy was then broken down into 
vdW and chemical contributions as described above for the single molecule case.
As a control, we also took a 28-molecule water cluster, moved it to the 
medium 11 \AA~pore, naming this model ``Model-28MP", 
fully relaxed the system (\fig{medium}), and broke down its energy
into components.  The results are shown in \tab{break}.  

 From \tab{break}, the primary cause for the high stability of water clusters in
\cubtc~is the {\em combination} of intracluster (hydrogen
bonding) interactions within each cluster and chemical interactions between the
cluster and the framework, particularly when each exposed Cu site inside the
large pore ``sees" a molecule belonging to a cluster.
The medium pores lack exposed Cu sites, and the Model 28-MP cluster in
this pore has almost zero chemical interaction with the framework.
Comparing Model 28-MP and Model 28, the intracluster interactions are stronger
in the medium pore, but the overall binding energy of 
28 water molecules is about 1.8 eV more stable inside the large pore,
Secondary influences on water cluster stability in \cubtc~include 
van der Waals interactions and secondary chemical interactions.
Secondary chemical interactions results from 
two sources: enhancement of \h2o-\h2o
binding under an electric field\cite{Choi06,Rai08,Mondal13}
(generated by the exposed Cu ions), and additional
H-bonding of \h2o~H to O in the framework itself, seen
in \fig{relaxed}(b) and (d).



All calculations were performed at zero temperature,
and do not include thermal motion. Given that water
is a liquid at room temperature, what can we say about 
the nature of the water in highly-hydrated \cubtc~
at room temperature?  Here, we refer to a recent
molecular dynamics simulation\cite{Haigis13} on ``MIL-53", a 
Cr-based MOF with a smaller unit cell than \cubtc.
They found, at room temperature, dynamic hydrogen
bonds, one $O_W$ (quasi)-statically bound to 
each exposed hydroxyl unit, and additional water molecules that move
fluidly.  This is akin to the picture discussed in
an NMR study on \cubtc,\cite{GulENoor13} where they conclude that there
are two types of water molecules: one type bound
to Cu, and the other fluid.  The experimental X-ray powder
diffraction refinement of \cubtc, showing 2.3 $O_W$ binding sites per 
Cu,\cite{Winnie14}
suggests that more water molecules are (quasi)-static
than just those immediately bound to the Cu, but that any
additional water molecules in the remaining pore space may be fluid
as their oxygen positions are not resolved.
Clearly, further theoretical and experimental studies 
on the dynamics of water in \cubtc~are needed.  
It would be useful to perform molecular dynamics studies of
\h2o~in \cubtc.  The relevant time scale for hydrogen
bonding rearrangements in \h2o~is of order
picoseconds (see {\it e.g} Ref.~\onlinecite{Keutsch01})
and measurements of liquid water diffusion rates\cite{Tanaka75} 
imply that \ow-\ow~rearrangements occur on the scale of tens of
picoseconds.  These time scales are too long for {\it ab initio}
molecular dynamics to be practical; classical molecular dynamics on 
a suitable force-field model are required.  On the experimental side, 
diffraction studies at low temperatures, where the water is completely
frozen, would be particularly useful.

Do our results provide guidance for the design of a MOF with weaker
water binding, one that might be more useful for \co2 capture
from a moist gas mixture?   In the following, we speculate that
the distance separating the metal ions in a MOF with exposed
metal ions is one factor that may affect the selectivity of
\h2o versus other sorbates.
For \cubtc~hydrated with one water
per Cu ion (\fig{relaxed}(a)), the closest \ow-\ow~distance
is 5.2~\AA, which coincides with the third-neighbor peak in 
the O-O pair distribution function of water ice.\cite{Geiger14}
On the other hand, an O-O distance of 5.8~\AA, is near a minimum 
of the O-O pair distribution function for both ice and liquid
water.\cite{Geiger14,Skinner13}  If one could synthesize a MOF with
open metal sites such that the equilibrium \ow-\ow~distance
between water molecule bound to adjacent metal ions were 5.8~\AA,
the formation of water clusters that bridge this unfavorable
\ow-\ow~distance might be inhibited, leading to a more favorable 
balance between \co2 and water sorption.

\section*{Conclusions}

Meta-GGA+U density functional theory calculations are
shown to reproduce the experimental structure of 
dry~\cubtc~and also to accurately give the interaction energies of
small water clusters.
Applying these calculations to models for hydrated \cubtc, we find
that hydrogen-bonded water clusters have their stability
enhanced primarily by interactions with the exposed metal ions, and
secondarily by van der Waals interactions, electric field enhancement of 
water-water bonding, and hydrogen bonding of water to framework oxygens.
This explains the great affinity of water for 
\cubtc~and related metal-organic frameworks.

\section*{Acknowledgements}

 We acknowledge W. Wong-Ng and L. Espinal 
for helpful discussions.


\begin{thebibliography}{48}
\bibitem[{foo({\natexlab{a}})}]{foot1}
\bibinfo{note}{Pore sizes based on placing spheres on each atom of the
  corresponding Van der Waals radius, and determining the largest sphere that
  can fit inside each pore, as well as the largest sphere that can pass through
  each window. Large pore sizes are bigger than reported in
  Ref.~\onlinecite{Vishny03} because the evaluation here is based on dry
  Cu-BTC.}
\bibitem[{dis()}]{disclaim}
\bibinfo{note}{Certain commercial software is identified in this paper to
  adequately describe the methodology used. Such identification does not imply
  recommendation or endorsement by the National Institute of Standards and
  Technology, nor does it imply that the software identified is necessarily the
  best available for the purpose.}
\bibitem[{foo({\natexlab{b}})}]{foot2}
\bibinfo{note}{The basis sets for the three GAUSSIAN results
  (Ref.~\onlinecite{Lee00}) that we averaged were: B3LYP/6-311++G**, MP2/DZP,
  and MP2/TZ2P++, each with 50\% BSSEC. The hexamer configuration was cyclic.}
\end{thebibliography}

\end{document}